\documentclass{article}

    \PassOptionsToPackage{numbers,compress}{natbib}

    \usepackage[preprint]{neurips_2024}

\usepackage[utf8]{inputenc} %
\usepackage[T1]{fontenc}    %
\usepackage{hyperref}       %
\usepackage{url}            %
\usepackage{booktabs}       %
\usepackage{amsfonts}       %
\usepackage{nicefrac}       %
\usepackage{microtype}      %
\usepackage{xcolor}         %
\usepackage{graphicx}       %
\usepackage{subcaption}     %
\usepackage{tikz}           %
\usepackage{multicol}       %
\usepackage{xspace}         %

\newcommand{\method}{\textsc{GTA}\xspace}

\title{Long-range gene expression prediction with token alignment of large language model}

\author{%
  Edouardo Honig\textsuperscript{1,2} \\ %
  \And
  Huixin Zhan\textsuperscript{2} \\
  \And
  Ying Nian Wu\textsuperscript{1} \\
  \And
  Zijun Frank Zhang\textsuperscript{2,3} \\
  \AND \vspace{-5mm}\\
  \textsuperscript{1}University of California, Los Angeles: Department of Statistics \& Data Science \\
  \textsuperscript{2}Cedars-Sinai Medical Center: Division of Artificial Intelligence in Medicine\\
  \textsuperscript{3}Cedars-Sinai Medical Center: Department of Computational Biomedicine\\
  \vspace{-10mm}
}

\begin{document}

\maketitle

\begin{abstract}
Gene expression is a cellular process that plays a fundamental role in human phenotypical variations and diseases. 
Despite advances of deep learning models for gene expression prediction, recent benchmarks have revealed their inability to learn distal regulatory grammar. 
Here, we address this challenge by leveraging a pretrained large language model to enhance gene expression prediction. 
We introduce \textbf{Genetic sequence Token Alignment (\method)}, which aligns genetic sequence features with natural language tokens, allowing for symbolic reasoning of genomic sequence features via the frozen language model. 
This cross-modal adaptation learns the regulatory grammar and allows us to further incorporate gene-specific human annotations as prompts, enabling in-context learning that is not possible with existing models. 
Trained on lymphoblastoid cells, \method was evaluated on cells from the Geuvadis consortium and outperforms state-of-the-art models such as Enformer, achieving a Spearman correlation of 0.65, a 10\% improvement. 
Additionally, \method offers improved interpretation of long-range interactions through the identification of the most meaningful sections of the input genetic context.
\method represents a powerful and novel cross-modal approach to gene expression prediction by utilizing a pretrained language model, in a paradigm shift from conventional gene expression models trained only on sequence data. 
\end{abstract}

\vspace{-2mm}
\section{Introduction}
\vspace{-2mm}
Genetic information takes a colossal role in determining human diseases and characteristics, and gene expression is the first step in the conversion of genetic sequence to phenotype. 
Understanding the genetic sequence determinants of gene expression is a difficult task due to the diversity of gene transcripts, sequences, and a limited scientific understanding of genomics. 
As such, there is much interest in building predictive models of gene expression from genetic sequence to improve the understanding of gene expression through \textit{in silico} experiments, which are often magnitudes cheaper than human studies. 
While prior work \cite{zhou2018deep, kelley2018sequential, kelley2020cross, agarwal2020predicting, avsec2021effective, sokolova2023atlas} has improved prediction accuracy, flaws in model extrapolation to unseen data \cite{huang2023personal, sasse2023benchmarking} and modeling of distal effects \cite{karollus2023current}. 
Therefore, predicting gene expression from sequence remains an open problem, the solving of which may lead to pivotal discoveries related to improving human healthcare. 

Gene expression begins at a transcription start site (TSS) which lies within a promoter, a regulatory element that is located immediately upstream of a gene. 
Many regulatory elements affect the gene expression process, such as enhancers, silencers, insulators, transcription factor (TF) binding sites, and 5 prime (5') or 3 prime (3') untranslated regions (UTR). 
While proximal elements such as promoters, TF binding sites, and 5' UTR tend to have increased impact on gene expression regulation compared to distal elements (enhancers, silencers, insulators), distal regulatory elements are demonstrated to have essential roles. 
For instance, the GTEx project \cite{gtex2020gtex} analyzes the expression quantitative trait loci (eQTL), the genetic variations that regulate target gene expression, in a 1 million basepairs (bp) window centered at TSS. 
These eQTLs are enriched in the various disease heritability, highlighting their functional importance \cite{finucane2018heritability, hormozdiari2018leveraging}. 
On the molecular mechanisms, due to the physical, three-dimensional, nature of Deoxyribonucleic acid (DNA), regions that may be physically close can appear distant in a one-dimensional sequence representation of the genetic code, emphasizing the necessity of incorporating both proximal and distal effects into models of gene expression from sequence.

Deep learning methods have been extensively applied to model gene expression from DNA sequences; however, all existing methods have limited sequence context length, thus are lacking the power to model distal regulatory elements (see Fig. \ref{fig:context_comparison}).
While initially deep convolutional neural networks \cite{lecun1989backpropagation} were used to incorporate information over long-ranges such as one-dimensional text sequences, recurrent and long short-term neural networks \cite{hochreiter1997long} became popular for auto-regressive sequence modeling. 
After its introduction, the Transformer \cite{vaswani2017attention} became widely adopted as a generally superior for modeling sequential data. 
Models that predict gene expression from sequence have followed a similar trend, from deep convolutional models \cite{zhou2018deep, kelley2018sequential, kelley2020cross, agarwal2020predicting, sokolova2023atlas} to Transformer-based models \cite{avsec2021effective}. 
However, the scale of auto-regressive language models has increased and eclipsed that of the most recent gene expression prediction model, Enformer \cite{avsec2021effective}. 
In-part due to such large language models being expensive to train and their emergent capabilities, there has been increased focus on applying pretrained large language models to multiple tasks \cite{brown2020language, raffel2020exploring} with transfer learning as general-purpose tools. 
Prior work \cite{lu2022frozen, jin2023time} has demonstrated the effectiveness of leveraging pretrained language models for non-language tasks via fine-tuning or cross-modal adaptation. 

We introduce \textbf{Genetic sequence Token Alignment (\method)}, a method that models sequence context up to 1 million bp for state-of-the-art gene expression prediction accuracy. This represents a five-fold increase of context length compared to the status quo. \method leverages frozen pretrained large language models to gene expression prediction from sequence by adapting genetic sequence class features to language model input embeddings. 
By aligning genetic sequence features to language model tokens, we can effectively model both proximal and distal effects across long sequences using a large language model, outperforming existing methods with a minimal amount of learnable parameters, avoiding further training of the backbone language model itself. 
To improve flexibility and minimize training cost, we use a pretrained genetic sequence model to extract the features in bins that we use as input for the token alignment process. 
This underlies the useful benefit of making it simple to increase or decrease the length of genetic context that is input to our model, since feature extraction is agnostic to location in the genome. 
Additionally, the usage of a pretrained language model begets the usage of in-context learning via natural language prompting, which is novel for the task of gene expression prediction. 
\method offers improved performance, flexibility, and the additional ability to incorporate human knowledge compared to prior gene expression prediction models.

\vspace{-2mm}
\section{Related Work}
\vspace{-2mm}
\begin{figure}[h!]
    \centering
    \includegraphics[width=0.95\textwidth]{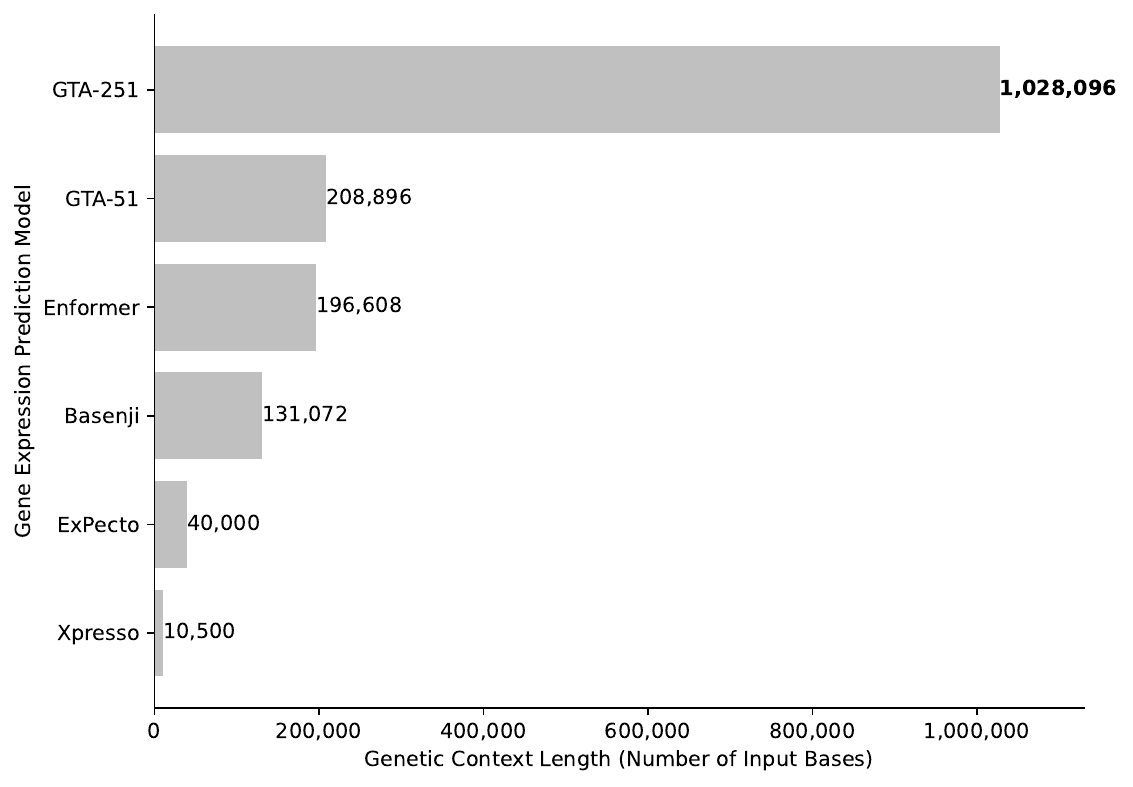}
    \caption{Comparison of input genetic sequence length for gene expression prediction models. \method enables a flexible range of input lengths, and we train models with input context from 200-1,000 kb.}
    \label{fig:context_comparison}
\end{figure}

\textbf{Gene Expression Prediction from Sequence}. 
Early work on gene expression prediction from sequence data used probabilistic approaches including Bayesian networks and Bayes classifiers and less than 1,000 base pairs of input sequence data \cite{beer2004predicting, yuan2007predicting}. 
Recent deep learning approaches have primarily focused on increasing the size of the genetic context to incorporate longer-range interactions. 
Initial approaches \cite{kelley2018sequential, zhou2018deep, agarwal2020predicting} used deep convolutional neural networks (CNNs) \cite{lecun1989backpropagation} to model long-range interactions across sequence. 
ExPecto \cite{zhou2018deep} uses 40 kilobases (kb) as input for a CNN that predicts 2,002 regulatory features, which are transformed and used for gene expression prediction via a regularized linear model. 
Xpresso \cite{agarwal2020predicting} uses 10 kilobases (kb) as input to a CNN that predicts gene expression directly from sequence. 
In contrast, Basenji \cite{kelley2018sequential} uses 131 kb for their CNN-based approach which predicts epigenetic and transcriptional profiles from sequence, from which a prediction for gene expression can be derived. 
A hybrid convolutional and Transformer-based \cite{vaswani2017attention} approach in Enformer \cite{avsec2021effective} builds upon Basenji by replacing dilated convolutions with transformer layers and uses an extended 200 kb context, improving performance over the purely convolutional prior work and emphasizing the importance of modeling long-range interactions. 
We train \method with a much larger context size than previous work, and train models that accept up to 1,000 kb as input (\autoref{fig:context_comparison}), with the ability to further increase the size of the context with minimal increase in number of model parameters. 
To do this, we follow an approach similar to ExPecto by using Sei \cite{chen2022sequence} to extract genetic sequence features we use as input for gene expression prediction with our method. 

\textbf{Transfer Learning and Model Reprogramming}. 
Transfer learning is a method in which a pre-initialized model is further trained on some new task or data \cite{hinton2006fast}. 
Works that consider using activations from trained networks such at DeCAF \cite{donahue2014decaf} are similar to ExPecto \cite{zhou2018deep} which uses the final output predictions of a pretrained network for a separate task. 
LoRA \cite{hu2021lora} trains adapters for a model's layers that allow fine-tuning a model without adjusting its original parameters. 
Model reprogramming \cite{chen2024model, jin2023time} is similar to transfer learning and LoRA, except it specifies training only additional input and output layers and leaving the pretrained model completely unaltered. 
Time-LLM \cite{jin2023time} exemplifies the reprogramming approach, by showcasing results on time series forecasting. 
\method similarly uses a reprogramming approach for gene expression prediction, which can incorporate more domain-specific information and better leverage the model reprogramming approach.

\vspace{-2mm}
\section{Methods}
\begin{figure}[ht]
    \centering
    \includegraphics[width=0.95\textwidth]{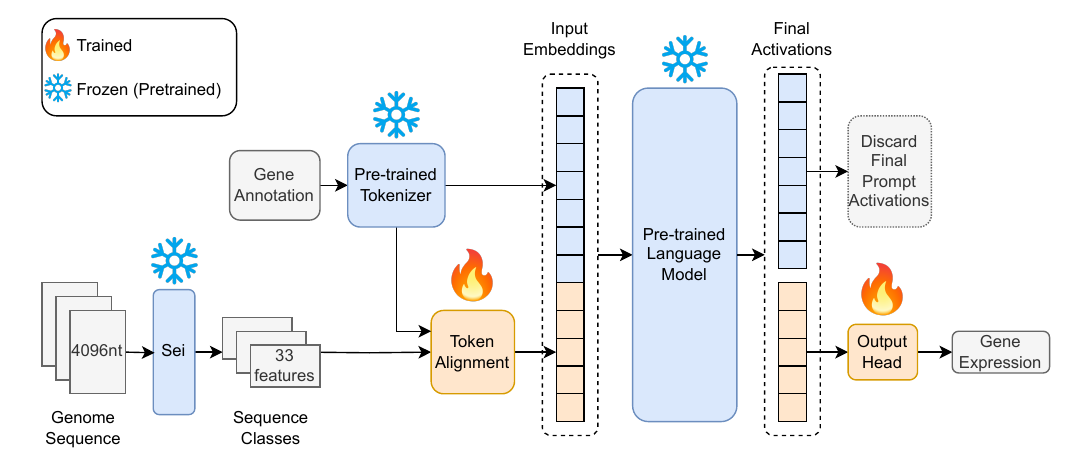}
    \caption{Overview of \method.}
    \label{fig:overview}
\end{figure}

\vspace{-2mm}
Our goal is to model gene expression $\mathbf{y} \in \mathbb{R}^1$ from sequence, which we achieve by aligning the representations of a sequence using a pretrained genomic sequence embedding model with tokens from a pretrained language model. 
An overview of our method can be seen in \autoref{fig:overview}. 

\textbf{Obtaining local epigenetic representations}. 
While much previous work using deep learning for gene expression prediction focuses on training a single model directly from one-hot encoded nucleotides that make up a genetic sequence \cite{kelley2018sequential, agarwal2020predicting, avsec2021effective}, we argue that it is difficult to learn feature extractors as the input genetic context length increases.
Additionally, there already exist adequate models for extracting useful features from genetic sequences, so it is unnecessary to train a model directly on one-hot encoded nucleotides. 
We use Sei \cite{chen2022sequence} to extract meaningful genetic features.
Sei is trained on 4,096 nucleotide input sequences to predict 21,907 chromatin profiles, which can be compressed into 40 regulatory activities (sequence classes), which were identified by clustering the predictions from 30 million genetic sequences tiling the human genome. 
The sequence classes obtained from Sei are annotated with names related to various functions, such as the brain, enhancers, promoters, as well as "low signal", which denotes "low enrichment in
the measured histone marks from the training of Sei" \cite{chen2022sequence}. 
We use Sei to obtain representations of non-overlapping subsections of a genetic sequence, dropping all "low signal" sequence classes, resulting in input data $\mathbf{X} \in \mathbb{R}^{N \times 33}$, where $N$ is defined as the number of non-overlapping 4,096 nucleotide bins that span the input genetic sequence for our model. 
Unless otherwise stated, we use $N=251$ in our experiments, which corresponds to a roughly one million nucleotide context length and is five times wider than the input receptive field used in Enformer \cite{avsec2021effective}. 

\textbf{Alignment of genetic features to language model tokens}. 
In order to leverage a pretrained language model for gene expression prediction without computationally expensive fine-tuning, we must manipulate our input features such that they can be processed akin to tokens in the language model's vocabulary. 
A naive mapping from the input feature space to the language model's embedding space is unlikely to be useful, since this ignores the token embeddings learned during the language model's training process. 
Instead, we desire an alignment procedure that transforms our input data with respect to the language model's vocabulary.   
Following \cite{jin2023time}, we build a learnable set of text prototypes for aligning our non-natural language input data. 
The text prototypes are linear combinations of the entire language model's learned token embeddings, since it is likely that only a small subset of the tokens in the language model's vocabulary will be related to our input data. 
Unlike \cite{jin2023time} where there is no prior knowledge regarding the relevancy of source tokens, our input features are named sequence classes. 
Because we have natural language descriptions for our features, we can directly select tokens in the vocabulary defined by the names of our input features as text prototypes as well. 
We therefore include a mapping from only the tokens from the sequence class names as frozen text prototypes in addition to the mapping from the entire vocabulary. 
In our experiments, a total of 100 text prototypes are used: 50 linear combinations of the entire vocabulary, and 50 linear combinations of the sequence class names. 
Finally, to align our input data with these text prototypes, we use neural cross-attention \cite{vaswani2017attention}. 
For a language model with embedding dimension size $D$, a number of heads $h$ for the alignment step with key dimension $d_k$, we have $d=\lfloor\frac{d_k}{h}\rfloor$, as in \cite{jin2023time}. 
The input data is projected into queries $Q \in \mathbb{R}^{N \times d}$ for the text prototypes' projected keys and values $\{K,V\} \in \mathbb{R}^{P \times d}$, where $P$ is the number of text prototypes. 
The attention $\mathbf{A}_i \in \mathbb{R}^{N \times d}$ for each head $i \in \{1,\ldots,h\}$ is calculated as:
\begin{equation}
    \mathbf{A}_i = {\rm Attention}(\mathbf{Q}_i, \mathbf{K}_i, \mathbf{V}_i) = {\rm softmax}(\frac{\mathbf{Q}_i \mathbf{K}^{\top}_i}{\sqrt{d_k}})\mathbf{V}_i
\end{equation}

and the attention $\mathbf{A} \in \mathbb{R}^{N \times d_k}$ is aggregated through concatenation then linearly mapped to the language model's embedding dimension $D$, such that the input sequence features have been aligned with the text prototypes and can be treated as tokens in the pretrained language model. 
We use $h=8, d_k=32$, and $D=4{,}096$ for Llama3-8B throughout all experiments.

\begin{figure}[ht]
    \centering
    \includegraphics[width=0.95\textwidth]{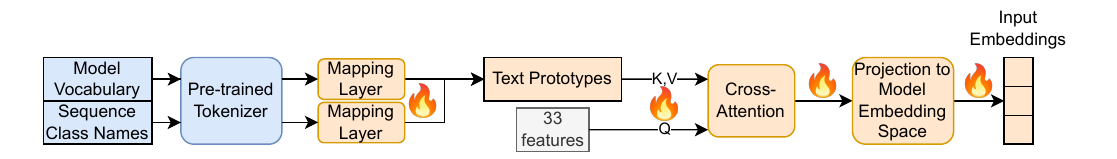}
    \caption{Detailed view of token alignment procedure.}
    \label{fig:token_alignment}
\end{figure}

\textbf{Incorporating context through gene annotations}. 
A common method to improve language model performance at inference time is through the incorporation of in-context information.
While it is difficult to incorporate chain-of-thought prompting \cite{wei2022chain} with our method, we can easily incorporate additional information through standard prompting. 
Since the human genome has been fully sequenced \cite{altemose2022complete}, we choose to add additional context on the genes whose data are being passed into the language model, in the form of annotations queried from the National Center for Biotechnology Information (NCBI) Gene service \cite{NCBI_Gene} through Entrez \cite{maglott2005entrez} via Biopython \cite{cock2009biopython}.
As scientific understanding of genetics advances and is included in language model training data, we can increasingly include more detailed information to improve the learned token alignment of genetic sequence features and text prototypes. \autoref{fig:gene_annotation} shows the prompt used for the gene A1BG, where the non-bold text besides the task description is gene-dependent.

\begin{figure}[h]
\centering
\begin{tikzpicture}
  \node[draw, rectangle, rounded corners, minimum width=10cm, minimum height=3cm, fill=blue!10, text width=5in, align=left] {
    \textbf{\# Task}: Gene Expression Prediction on GM12878 lymphoblastoid cells.\\
    \textbf{\# Official Gene Symbol}: A1BG\\
    \textbf{\# Official Gene Full Name}: alpha-1-B glycoprotein\\
    \textbf{\# Other Gene Aliases}: A1B, ABG, GAB, HYST2477\\
    \textbf{\# Other Gene Designations}: alpha-1B-glycoprotein|HEL-S-163pA|epididymis secretory sperm binding protein Li 163pA\\
    \textbf{\# Gene NCBI Summary Description}: The protein encoded by this gene is a plasma glycoprotein of unknown function. The protein shows sequence similarity to the variable regions of some immunoglobulin supergene family member proteins. [provided by RefSeq, Jul 2008]\\
    \textbf{\# Data}: 
  };
\end{tikzpicture}
\caption{NCBI Gene Annotation adopted from \url{https://www.ncbi.nlm.nih.gov/gene/1}.}
\label{fig:gene_annotation}
\end{figure}

\textbf{Selecting final activations for regression}. 
Our main goal of re-purposing a language model for gene expression prediction is to better model long-range interactions across the genetic sequence context. 
As such, while many methods that re-purpose language models for regression or classification use only the final activations from a single token \cite{devlin2018bert, dosovitskiy2021an}, we hypothesize that regressing gene expression on multiple the final activations of multiple aligned tokens will improve predictive performance. 
As such, we use the logits of a linear layer to act as a \textbf{selector network} for the final activations to feed into our final linear output layer to predict gene expression. 
Unlike Enformer, which extracts features from the input DNA sequence in 128-base bins and consistently crops out regions on both ends of the sequence before prediction, we learn to flexibly select the areas to include or exclude from the entirety of the input data. 
Note that the top-$k$ final activations are selected from our aligned data tokens only, excluding the gene annotation prompt. 
We use $k=25$ in our experiments unless otherwise noted. 

\vspace{-2mm}
\section{Results}
\vspace{-2mm}
\method outperforms existing models for gene expression prediction from sequence on a task outlined in \cite{huang2023personal}, where the goal is to predict median gene expression of measurements from 421 individuals from the Geuvadis consortium \cite{lappalainen2013transcriptome} using the reference genome.
We hold out the same test gene set as in \cite{huang2023personal} and use the remaining gene set for model training.
We use Adam \cite{kingma2014adam} to optimize \method for gene expression prediction by minimizing the mean-squared error between predicted and observed values. 
Llama3-8B \cite{llama3modelcard2024} is the frozen pretrained backbone in all experiments with a reduced sequence length of 2,048 via rotary positional embedding (RoPE) \cite{su2021roformer} using UnslothAI \cite{unsloth_ai} to improve computational efficiency. 
Training was conducted using a single Nvidia A100-80GB GPU using PyTorch \cite{paszke2019pytorch} Lightning \cite{falcon2019pytorch}.

\subsection{Baseline Comparison}

We compare \method with Xpresso, ExPecto, Basenji, and Enformer on gene expression prediction of GM12878 Lymphoblastoid cells. 
We train and validate our model on a set of genes taken as the intersection of those used for the creation of the baseline models, and test performance by holding-out the set of genes used by \cite{huang2023personal}. 
After filtering out genes in the evaluation data as well as genes for which no Matched Annotation from NCBI and EMBL-EBI (MANE) \cite{morales2022joint} transcription start site (TSS) was found, we split the 15,087 genes into 80\% training and 20\% validation splits, and maintain 2,850 genes for evaluation with existing MANE TSS. 
To obtain the Sei sequence features, we extracted sequences of $251 \times 4{,}096 = 1{,}028{,}096$ bases centered on each gene's MANE TSS using the GRCh38 reference genome, padding when necessary with a placeholder base "N", separated into 251 non-overlapping bins, each containing 4,096 bases, which were fed into Sei to obtain sequence class features. 
The "low signal" Sei sequence classes were then removed, leaving 33 sequence classes and our input data as $\mathbf{X} \in \mathbb{R}^{N \times 33}$. 
To incorporate gene annotations as prompts, we tokenize each annotation and truncate it to $2{,}048 - N$ tokens so that the final $N$ tokens are the sequence class features.

\begin{table}[htp]
  \caption{Performance Evaluation}
  \label{tab:eval-spearman}
  \centering
  \begin{tabular}{lr}
    \toprule
    Model     & Spearman Correlation ($\uparrow$) \\
    \midrule
    Xpresso  & $0.2984 \pm 0.0351$ \\
    ExPecto  & $0.5170 \pm 0.0304$ \\
    Basenji  & $0.5234 \pm 0.0313$ \\
    Enformer & $0.5814 \pm 0.0293$ \\
    \midrule
    \method  ($N=51,  \approx 200$ kb input) & $0.6259 \pm 0.0256$ \\ %
    \method  ($N=101, \approx 400$ kb input) & $0.6167 \pm 0.0251$ \\ %
    \method  ($N=151, \approx 600$ kb input) & $0.6078 \pm 0.0250$ \\ %
    \method  ($N=201, \approx 800$ kb input) & $0.6227 \pm 0.0246$ \\ %
    \textbf{\method ($\mathbf{N=251, \approx 1000}$ kb input)} & $\mathbf{0.6527 \pm 0.0241}$ \\ %
    \bottomrule
  \end{tabular}
\end{table}

We evaluate the Spearman correlation of the predictions of each model with the observed values. 
Error bars are calculated as twice the standard deviation of 5,000 sample bootstraps.
As can be seen in \autoref{tab:eval-spearman}, \method with a variety of input context lengths outperforms prior models, and the model with the largest input context obtains the best performance. 
While \method is trained with the evaluation data completely held out, there is an overlap in the training data of prior models and the evaluation data, meaning there could potentially be data leakage in the baseline predictions (but not in \method's).

\begin{figure}[h]
    \centering
    \begin{subfigure}[b]{0.475\textwidth}
        \centering
        \includegraphics[height=2.5in]{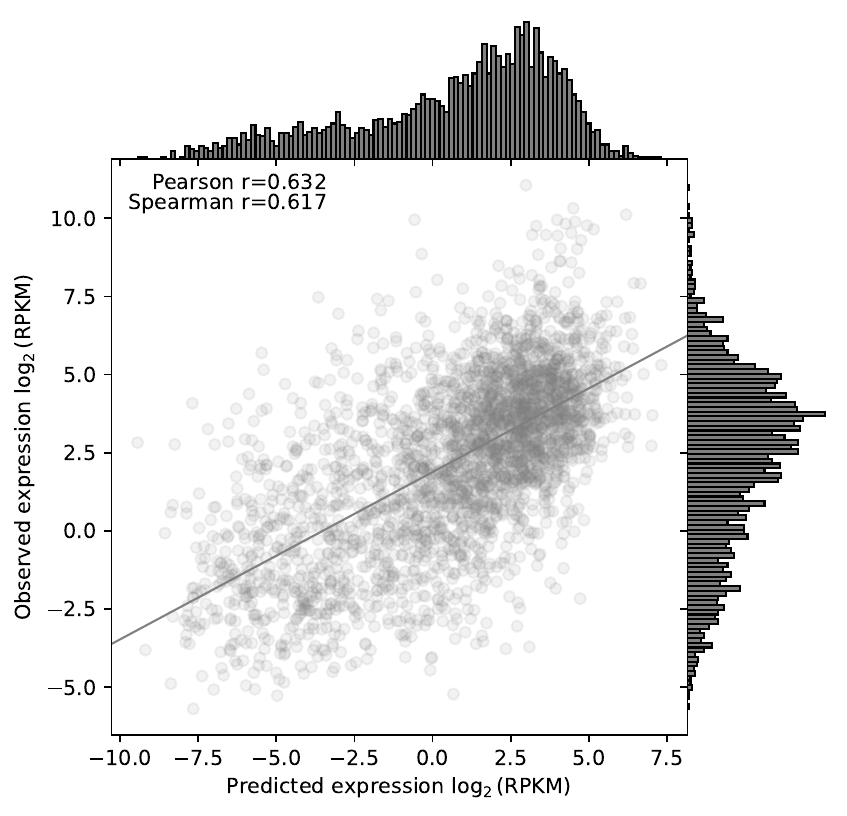}
        \caption{\method (200 kb input context)}
        \label{fig:eval-scatter200}
    \end{subfigure}
    \begin{subfigure}[b]{0.475\textwidth}
        \centering
        \includegraphics[height=2.5in]{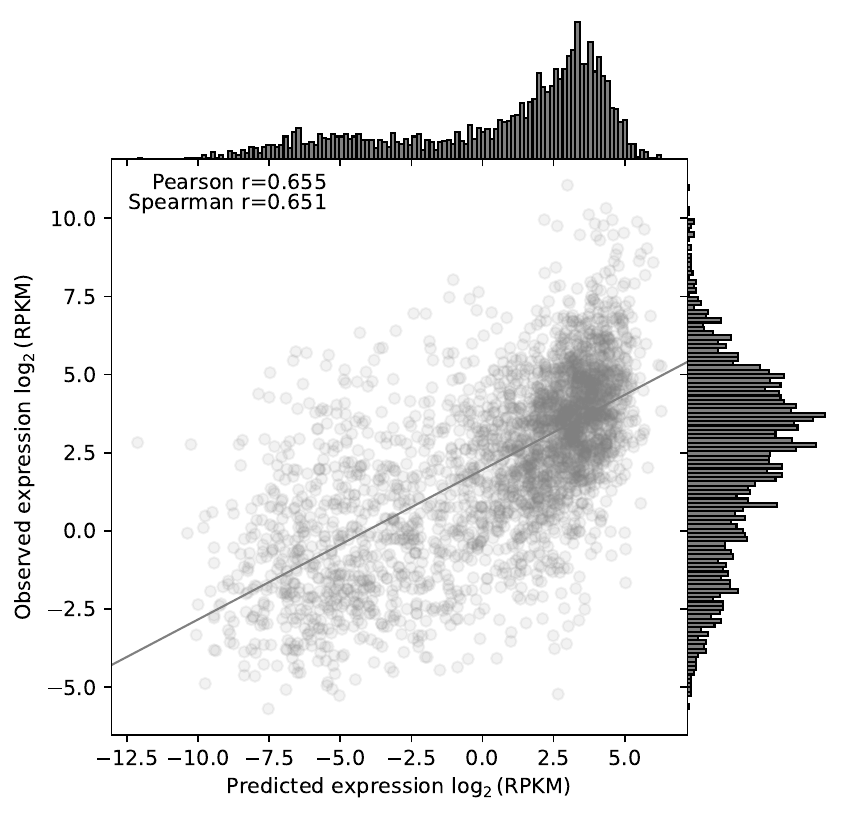}
        \caption{\method (1000 kb input context)}
        \label{fig:eval-scatter1000}
    \end{subfigure}
    \caption{\method predictions on the evaluation data.}
\end{figure}

\begin{figure}[h]
    \centering
    \includegraphics[height=3in]{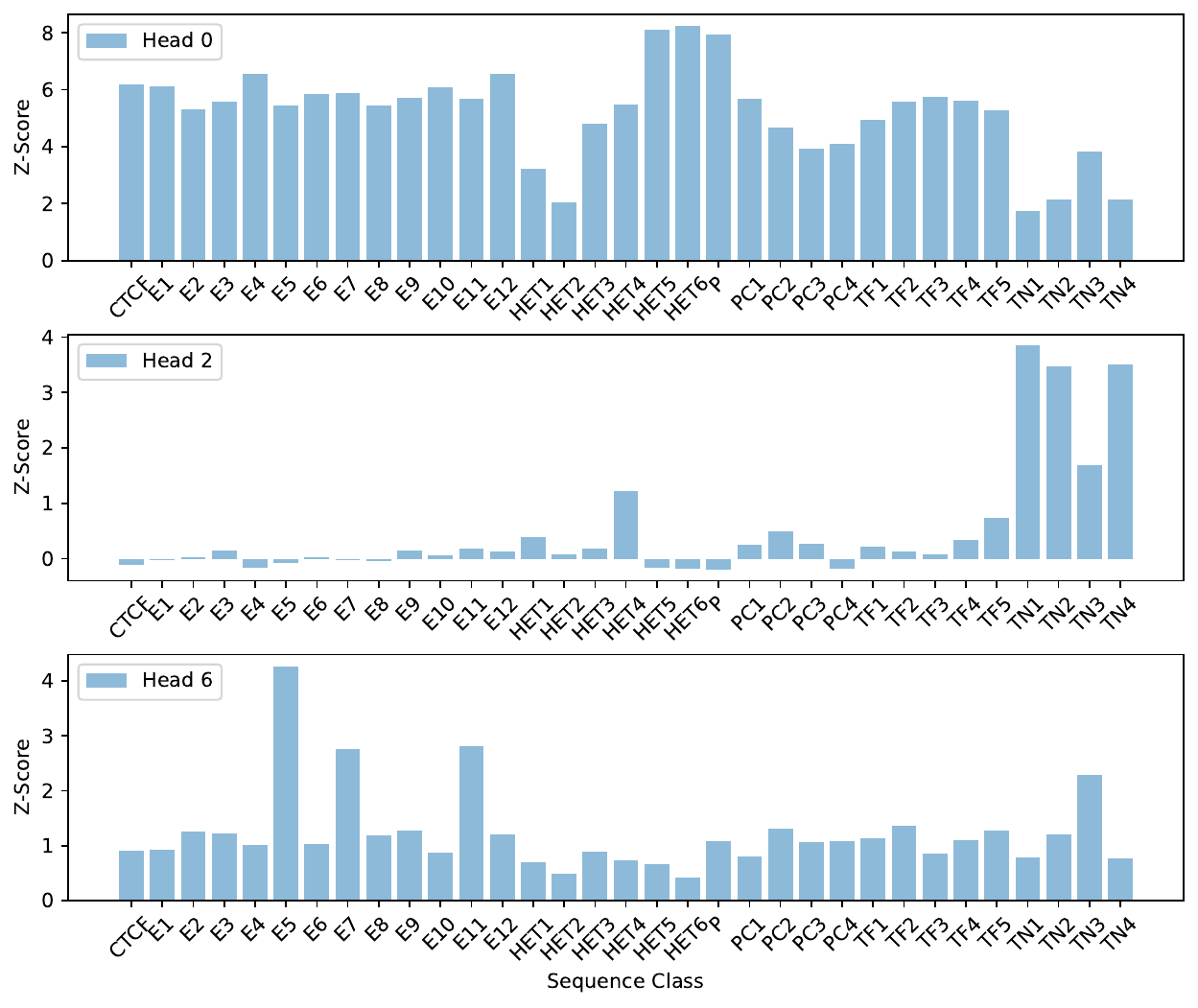}
    \caption{Z-scores of token alignment attention heads that capture the regulatory grammar.}
    \label{fig:xattn_heads}
    \vspace{-5mm}
\end{figure}

\subsection{Token Alignment}

A main goal of the token alignment procedure is to identify the regulatory grammar during the cross-modal adaptation via the attention mechanism. 
We identify the text prototypes with maximal cross-attention scores and select the input context bins with maximal attention scores for these text prototypes, for each attention head and each gene in the evaluation data. 
The features for these selected bins normalized with respect to the entire evaluation data for select attention heads are displayed in \autoref{fig:xattn_heads}. 

The cross-modal adaptation successfully learns biologically meaningful attention mechanism in a completely data-driven avenue. 
Notably, head 0 attends to all regulatory features with the exception of transcriptions (TN), which head 2 attends to almost exclusively. 
Attention head 6 attends mainly to B-cell, Monocyte, and T-cell enhancers (E5, E7, E11, respectively), which are related to immune cells. 
The remaining five attention heads learn similar feature importance as head 0, and a dictionary for the sequence class abbreviations and a visualization with all the heads can be found in \autoref{appendix:seqclassdefs} and \autoref{appendix:xattn_heads_all}.
Given our training dataset is GM12878 Lymphoblastoid cells, an immortalized cell-line derived by infecting peripheral blood lymphocytes with the Epstein-Barr virus, the fact that head 6 exclusively learns the circulating immune cell-specific enhancers underscores \method's ability to capture the distinct biological feature in this dataset.

\begin{figure}[h!]
    \centering
    \begin{subfigure}[b]{0.49\textwidth}
        \centering
        \includegraphics[height=2.3in]{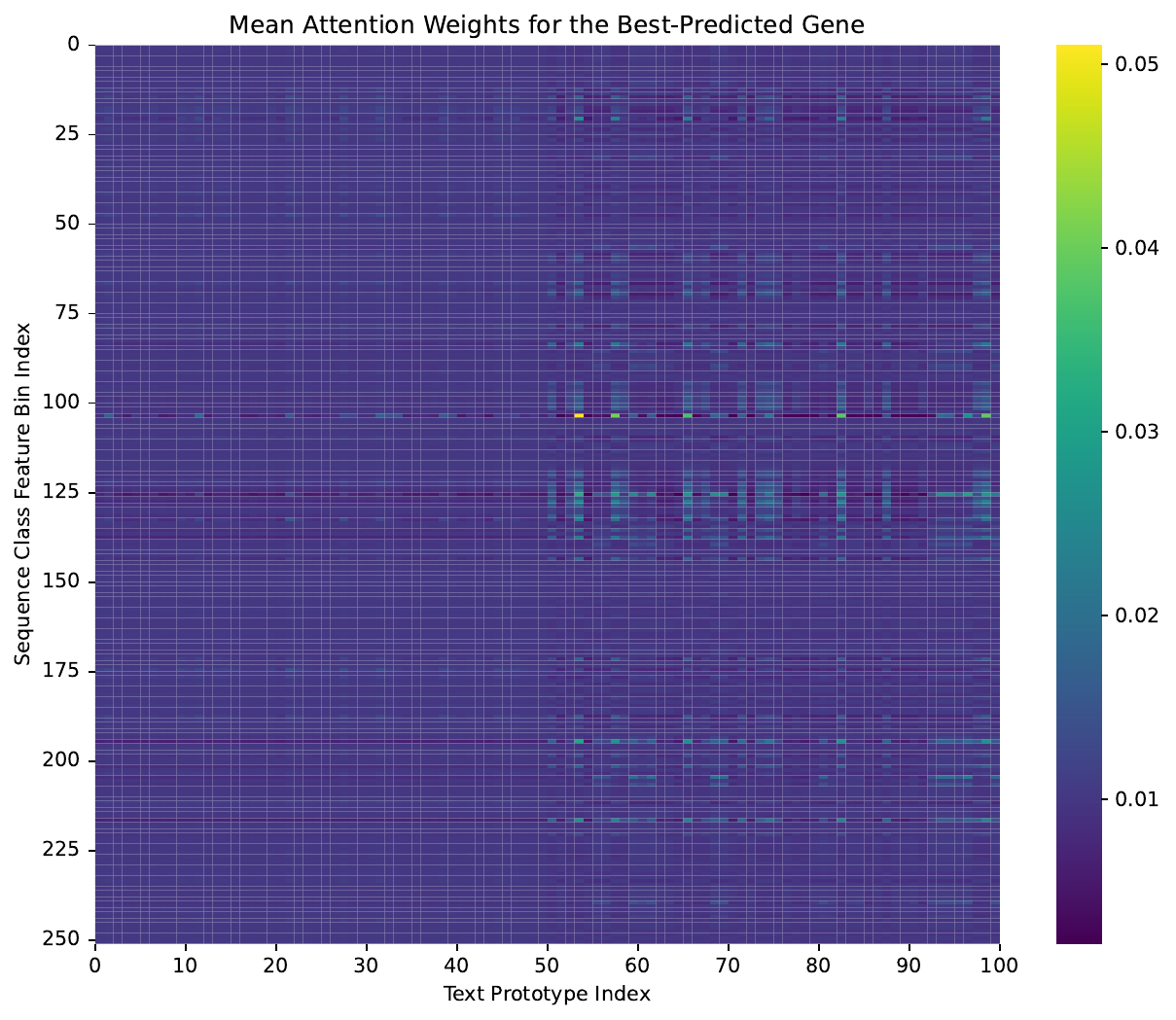}
        \caption{Best-predicted gene (${\rm residual}=0.0005$).}
        \label{fig:xattn_heatmap_best}
    \end{subfigure}
        \begin{subfigure}[b]{0.49\textwidth}
        \centering
        \includegraphics[height=2.3in]{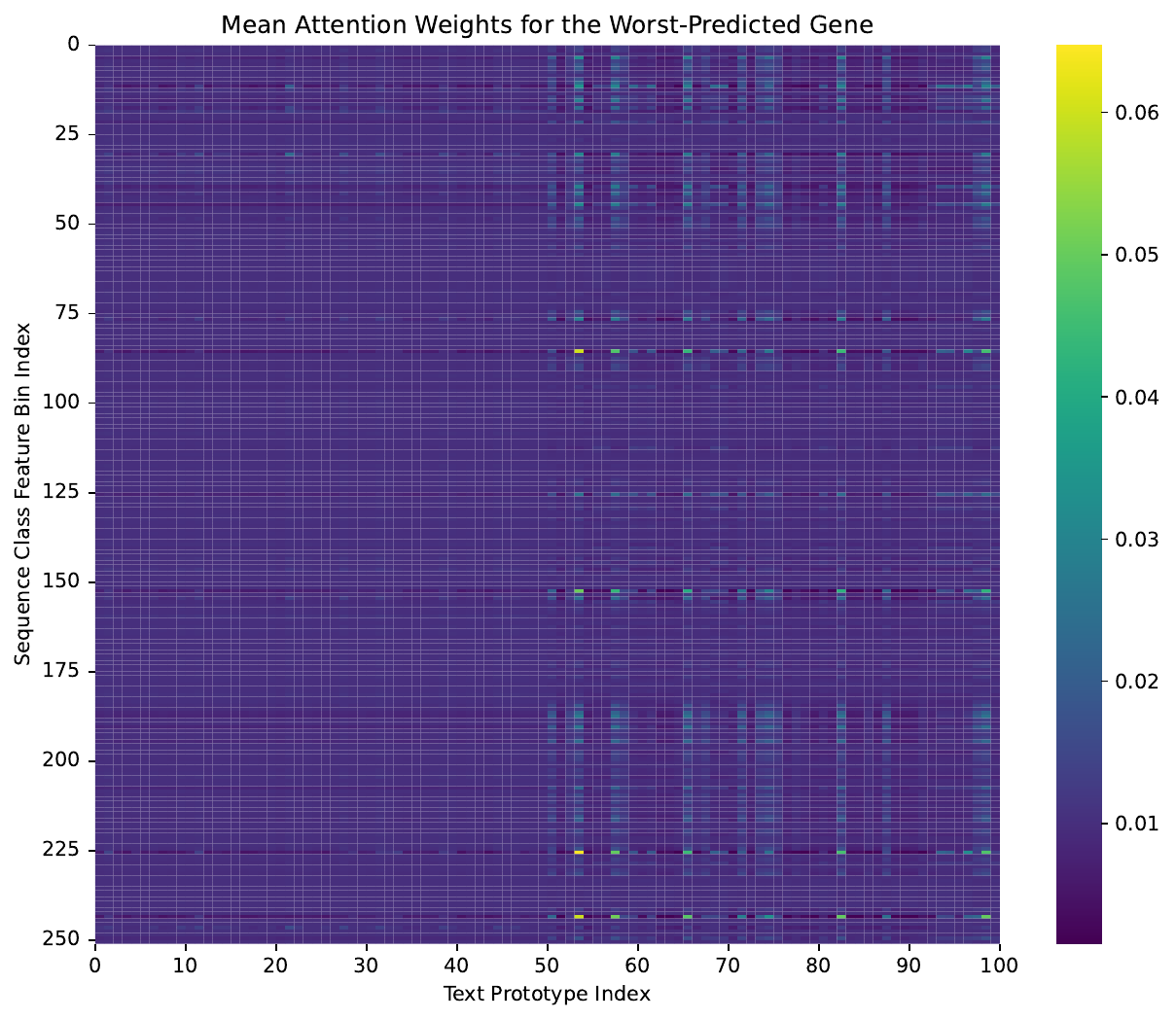}
        \caption{Worst-predicted gene (${\rm residual}=15.1$).}
        \label{fig:xattn_heatmap_worst}
    \end{subfigure}
    \caption{Cross-attention scores of all text prototypes and input data bins, averaged across heads. The first 50 prototypes are mapped from the entire vocabulary, while the final 50 prototypes are mapped from the sequence class names. Best and worst are determined via minimal and maximal residuals.}
    \label{fig:xattn_heatmaps}
\end{figure}

Next, we hypothesize that the text prototypes mapped from the genetic sequence class vocabulary will be more useful than those learned from the entire vocabulary. 
We visualize the averaged cross-attention scores across heads in the token alignment at a high level in \autoref{fig:xattn_heatmaps}, displaying scores for the two genes with the highest and lowest residuals respectively. 
The text prototypes mapped from the sequence class names (50-99) show much higher attention scores compared to those mapped from the entire model vocabulary, supporting the use of annotated genetic sequence classes over other sequence embedding methods.
While the text prototypes attend to similar parts of the genetic context, many sequence bins with greater attention scores exist close to the TSS (bin 125) in \autoref{fig:xattn_heatmap_best}, which is in-line with current scientific understanding of proximal effects. 
This is in contrast with \autoref{fig:xattn_heatmap_worst}, where the scores near the edges of the input context are much higher than those near the TSS.

\begin{figure}[ht]
    \centering
    \begin{subfigure}[b]{0.49\textwidth}
        \centering
        \includegraphics[width=2.7in]{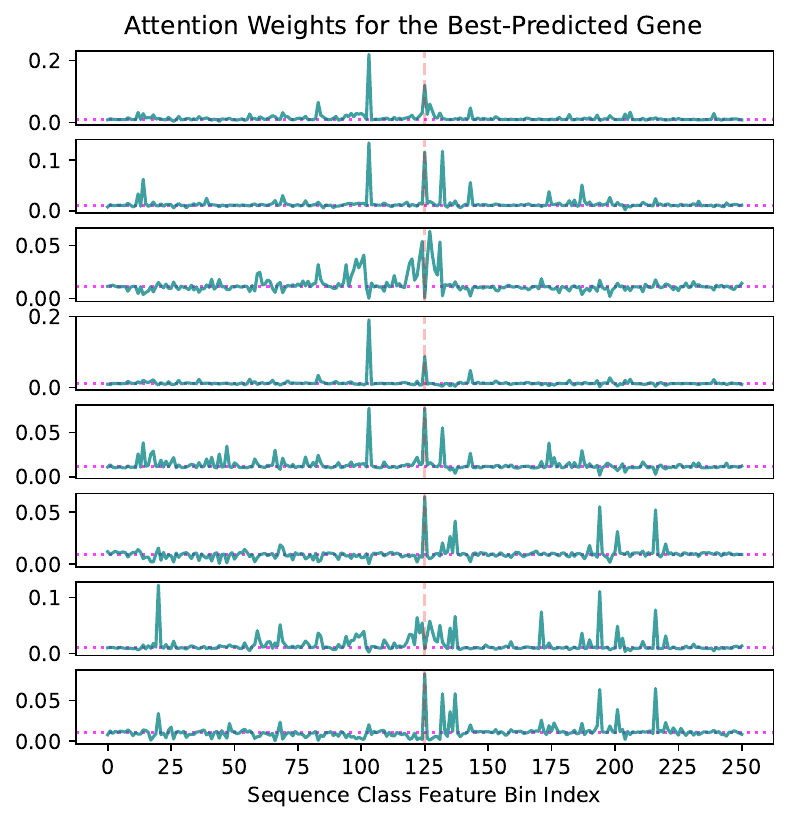}
        \caption{Best-predicted gene.}
        \label{fig:xattn_detail_best}
    \end{subfigure}
    \begin{subfigure}[b]{0.49\textwidth}
        \centering
        \includegraphics[width=2.7in]{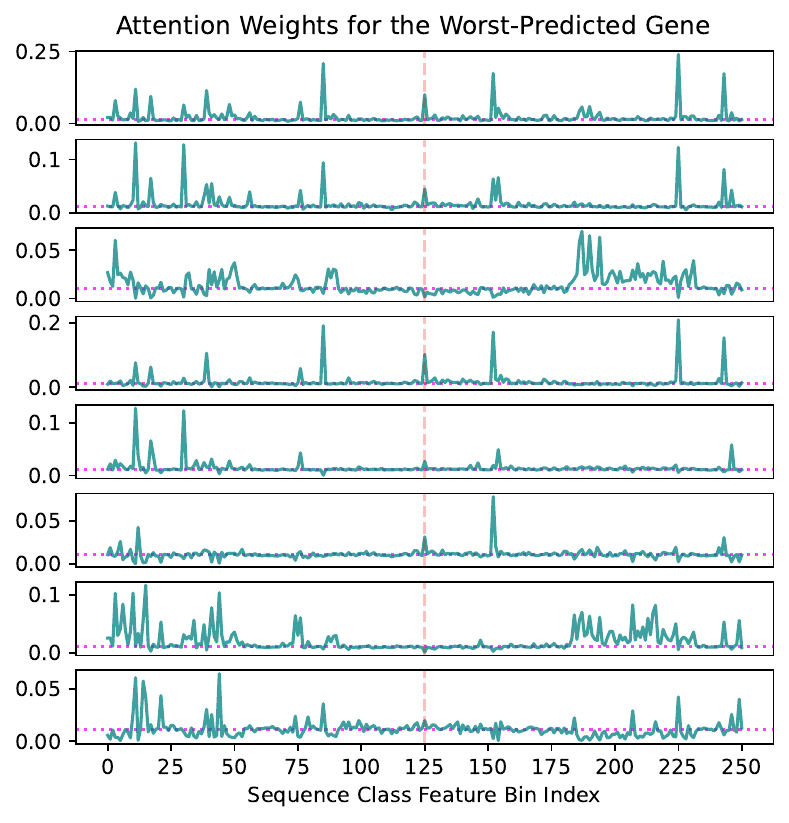}
        \caption{Worst-predicted gene.}
        \label{fig:xattn_detail_worst}
    \end{subfigure}
    \caption{Multi-head cross-attention for the text prototypes with the greatest average score per head. The TSS bin is indicated with the vertical line, and the horizontal line indicates the median score.}
    \label{fig:xattn_detail}
    \vspace{-2mm}
\end{figure}

\autoref{fig:xattn_detail} offers a more detailed view of the gene that was best predicted by our model. 
For the highest scoring text prototype, the TSS is the second most-attended to, and of the bins with the five greatest attention scores, three lie a far distance from the TSS, providing evidence that the token alignment process learns to model long-range distal effects. 
A contrasting comparison can be seen in the worst-predicted gene in \autoref{fig:xattn_detail_worst}, which shows lower attention scores for the TSS compared to the ends of the sequence. 
The high residual combined with this visualization suggests that the model is attending to the wrong areas of the genomic context for this specific gene.

In typical auto-regressive language modeling, only the final activation of the final token is used to predict the probability of the next token. 
However, due to the pretrained language model being frozen in \method, it is not necessarily optimal to adopt the same approach. 
While the backbone model will be useful for computation, we cannot expect all of the important information from the entire context to be aggregated in the final token without further training of the backbone model. 
While Time-LLM \cite{jin2023time} uses all final activations of the tokens input to the backbone model, this quickly becomes expensive as the number of input tokens and the hidden dimension of the backbone model increase. 
We therefore use the logits of a linear layer to select $k=25$ final activations to flatten as input for our output regression head, with the goal that this selection process will identify the most relevant tokens for gene expression prediction. 

\begin{figure}[ht]
    \centering
    \includegraphics[height=2.5in]{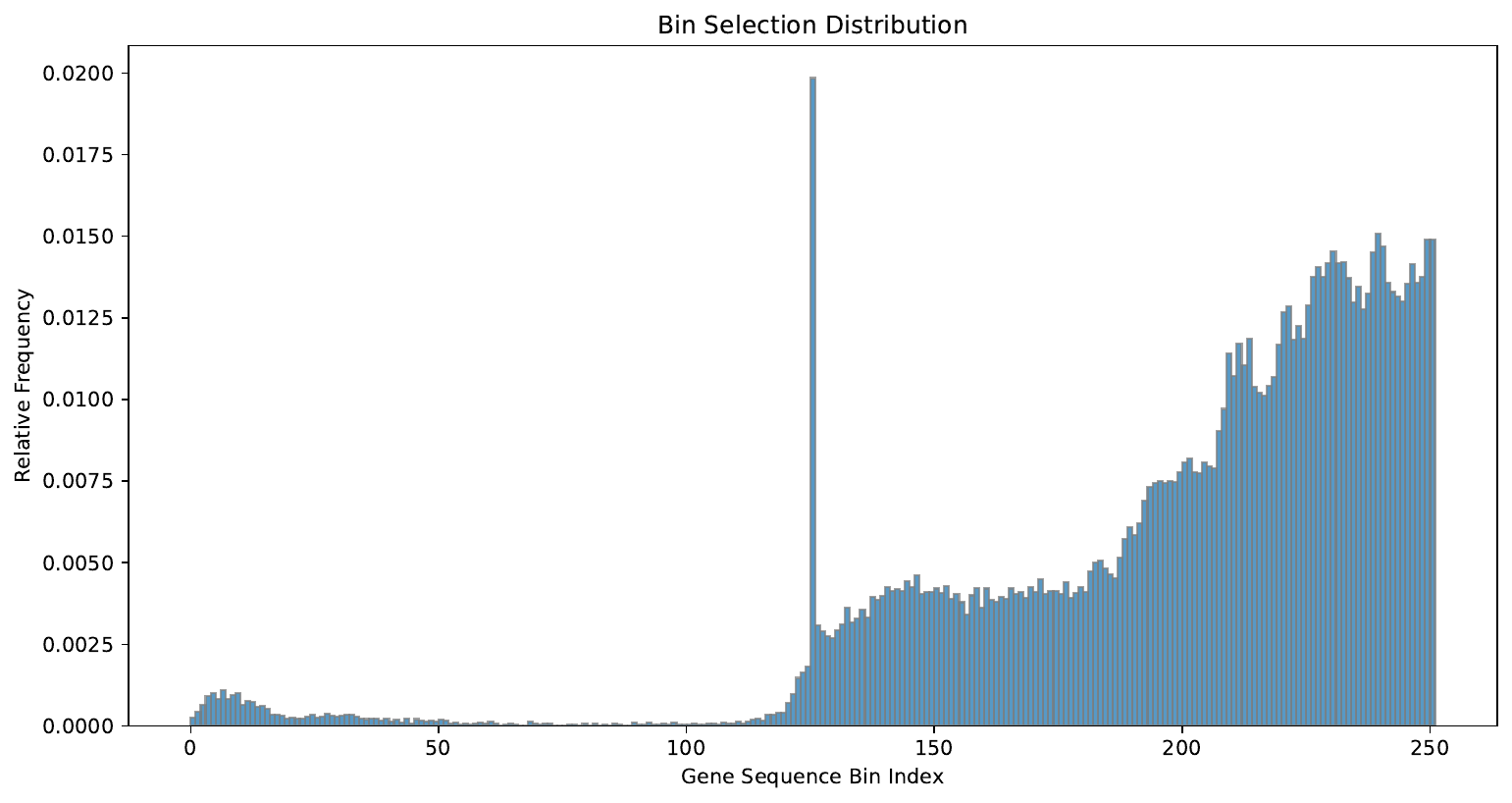}
    \caption{Frequency of bin selection for output regression head.}
    \label{fig:bin_selection}
\end{figure}

\autoref{fig:bin_selection} displays the distribution of selected output bins from the entire evaluation data set. 
Unsurprisingly, the TSS bin is the most frequently selected, although it is only selected for half of the genes in the evaluation set. 
Since we use a backbone model with causal attention, it is expected that earlier bins would be selected less, since they receive less attention than bins at the end of the sequence. 
This is potentially a limitation, since gene expression regulators can communicate in both directions on a genetic sequence. 
Using non-causal or full attention would alleviate this problem, but complicate the use of the most available and powerful (decoder-only) language models for \method. 
We leave it to future work to experiment with adjusting the attention mechanism in a decoder-only model to be non-causal.

\subsection{Ablation Study}

In the model reprogramming framework, it is not necessary and may not be optimal to use all of the parameters in a backbone model. 
Prior work \cite{openai2020imagegpt} has identified that the activations from a network that result in the best features may be activations from intermediate layers, rather than the final. 
Due to the backbone model being frozen, the activations fed to the output can be probed from any arbitrary layer from the backbone model, allowing for a reduction in total model parameters by removing unused layers. 
We detail the results of training multiple versions of \method with a varying subset of backbone layers in \autoref{tab:ablation-layers}, where increasing the number of transformer layers past 8 does not improve performance. 
These results informed the decision to train the \method models reported in \autoref{tab:eval-spearman} using only $\frac{8}{32}$ Llama3-8B transformer blocks, a large decrease in required computation. 

\begin{table}[htp]
  \caption{Performance Based on Backbone Size}
  \label{tab:ablation-layers}
  \centering
  \begin{tabular}{lr}
    \toprule
    Number of Transformer Blocks     & Spearman Correlation ($\uparrow$) \\
    \midrule
    2  & $0.5753 \pm 0.0268$ \\ %
    4  & $0.6067 \pm 0.0256$ \\ %
    \textbf{8}  & $\mathbf{0.6527 \pm 0.0241}$ \\ %
    16 & $0.5420 \pm 0.0272$ \\ %
    32 & $0.5913 \pm 0.0258$ \\ %
    \bottomrule
  \end{tabular}
\end{table}

\vspace{-2mm}
\section{Limitations}
\vspace{-2mm}
Our work assumes that the genetic sequence features learned by a pretrained model (in this work, Sei) adequately represent the genetic sequences that they embed. 
This compression is necessary for our method, which strongly outperforms prior work, but still potentially limiting, as \method does not learn from raw nucleotide data. 
We additionally require the features to have semantic meaning to learn improved text prototypes, which limits the choice of genomic feature extractor.

\vspace{-2mm}
\section{Conclusion}
\vspace{-2mm}
We introduce \method, a novel approach to gene expression prediction from sequence that aligns genetic sequence data to language tokens so as to leverage a frozen pretrained language model. 
\method outperforms prior gene expression models, easily scales to increasing genomic sequence context length, and is the first such model to explicitly incorporate human knowledge via gene annotations as prompts due to the use of the language model backbone. 
Through analysis of the cross-attention mechanism learned during token alignment, we demonstrate that the model learns aspects of the regulatory grammar through attention heads. 
The proper choice of inductive bias for the text prototypes via the source tokens improves performance compared to learning text prototypes as combinations of the entire model vocabulary. 
By demonstrating the effectiveness of the model reprogramming framework for gene expression, we motivate future work on incorporating data from various modalities to incorporate scientific knowledge and improve gene expression prediction. 

\begin{ack}
This work utilizes Llama3-8B, which is licensed under the Meta Llama 3 Community License.
\end{ack}

\bibliography{0_main.bib}

\newpage
\clearpage

\appendix
\section*{Appendix}
\section{Sei Sequence Classes Names}\label{appendix:seqclassdefs}

\begin{table}[!ht]
    \centering
    \begin{tabular}{llll}
    \hline
        Sequence class label & Sequence class name & Rank by size & Group \\
        \toprule
        PC1 & Polycomb / Heterochromatin & 0 & PC \\
        L1 & Low signal & 1 & L \\
        TN1 & Transcription & 2 & TN \\
        TN2 & Transcription & 3 & TN \\
        L2 & Low signal & 4 & L \\
        E1 & Stem cell & 5 & E \\
        E2 & Multi-tissue & 6 & E \\
        E3 & Brain / Melanocyte & 7 & E \\
        L3 & Low signal & 8 & L \\
        E4 & Multi-tissue & 9 & E \\
        TF1 & NANOG / FOXA1 & 10 & TF \\
        HET1 & Heterochromatin & 11 & HET \\
        E5 & B-cell-like & 12 & E \\
        E6 & Weak epithelial & 13 & E \\
        TF2 & CEBPB & 14 & TF \\
        PC2 & Weak Polycomb & 15 & PC \\
        E7 & Monocyte / Macrophage & 16 & E \\
        E8 & Weak multi-tissue & 17 & E \\
        L4 & Low signal & 18 & L \\
        TF3 & FOXA1 / AR / ESR1 & 19 & TF \\
        PC3 & Polycomb & 20 & PC \\
        TN3 & Transcription & 21 & TN \\
        L5 & Low signal & 22 & L \\
        HET2 & Heterochromatin & 23 & HET \\
        L6 & Low signal & 24 & L \\
        P & Promoter & 25 & P \\
        E9 & Liver / Intestine & 26 & E \\
        CTCF & CTCF-Cohesin & 27 & CTCF \\
        TN4 & Transcription & 28 & TN \\
        HET3 & Heterochromatin & 29 & HET \\
        E10 & Brain & 30 & E \\
        TF4 & OTX2 & 31 & TF \\
        HET4 & Heterochromatin & 32 & HET \\
        L7 & Low signal & 33 & L \\
        PC4 & Polycomb / Bivalent stem cell Enh & 34 & PC \\
        HET5 & Centromere & 35 & HET \\
        E11 & T-cell & 36 & E \\
        TF5 & AR & 37 & TF \\
        E12 & Erythroblast-like & 38 & E \\
        HET6 & Centromere & 39 & HET \\
        \bottomrule
    \end{tabular}
    \caption{Ranked Sei Sequence Classes}
    \label{tab:seqclass_names}
\end{table}

The Sei sequence classes are grouped into the following categories:

\begin{multicols}{2}
\begin{itemize}
    \item \textbf{CTCF}: CTCF-cohesin binding
    \item \textbf{E}: Enhancer
    \item \textbf{HET}: Heterochromatin
    \item \textbf{L}: Low Signal
    \item \textbf{P}: Promoter
    \item \textbf{PC}: Polycomb-repressed
    \item \textbf{TF}: TF binding
    \item \textbf{TN}: Transcription
\end{itemize}
\end{multicols}

Refer to \cite{chen2022sequence} for more details.

\section{Token Alignment Attention Heads}\label{appendix:xattn_heads_all}

\begin{figure}[h]
    \centering
    \includegraphics[width=0.95\textwidth]{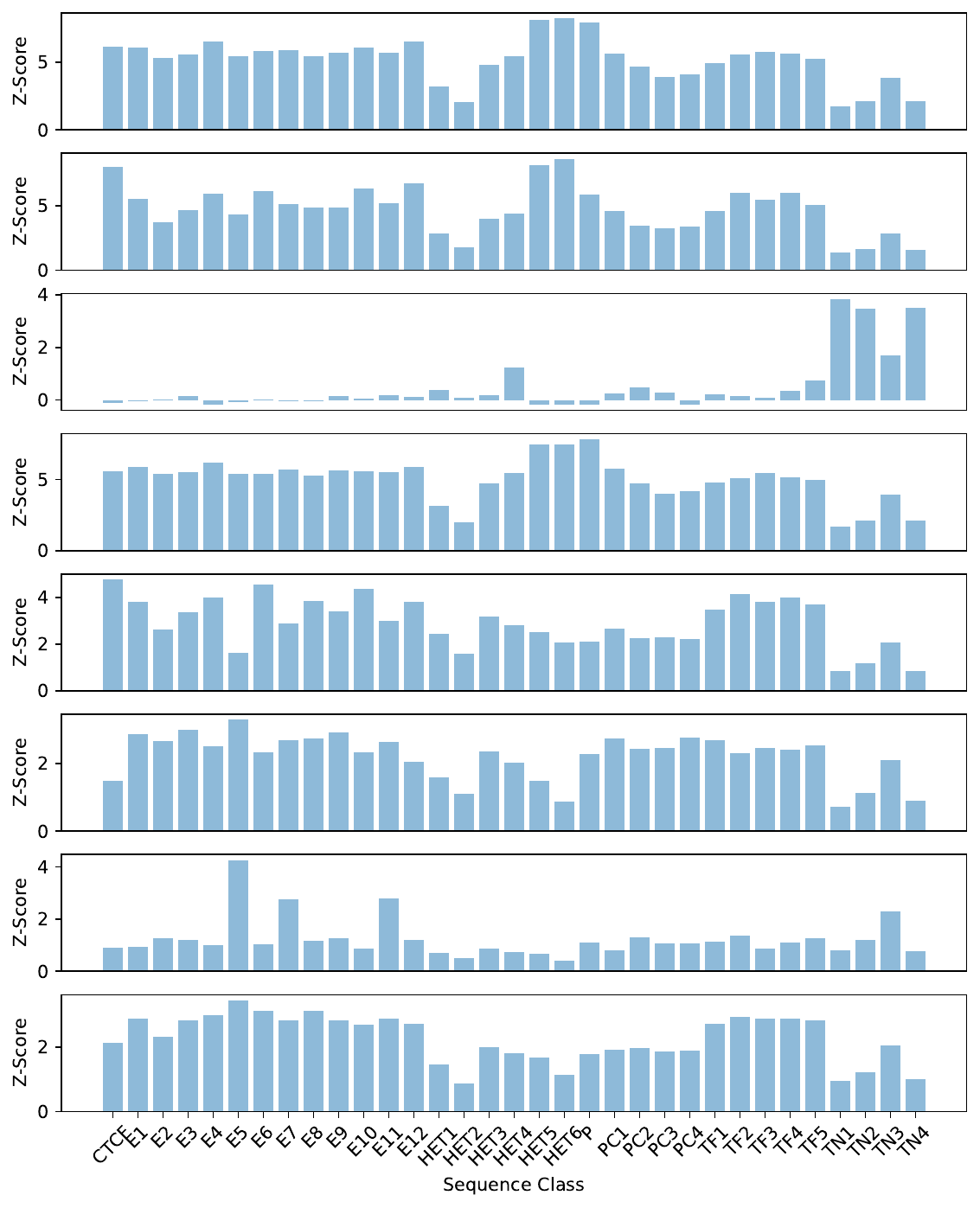}
    \caption{Z-scores of all token alignment attention heads including \autoref{fig:xattn_heads}, visualized for the evaluation data.}
    \label{fig:xattn_heads_all}
\end{figure}

\end{document}